# NORTH–SOUTH ASYMMETRY OF DIFFERENT SOLAR ACTIVITY FEATURES DURING SOLAR CYCLE 23

Neeraj Singh Bankoti[1]*, Navin Chandra Joshi[1], Seema Pande[2], Bimal Pande[1], Kavita Pandey[1]

[1]Department of Physics, D. S. B. Campus, Kumaun University, Nainital, Uttarakhand, India

[2]Depertment of Physics, M.B.P.G. College, Haldwani, Kumaun University, Nainital, Uttarakhand, India

*E-mail address.nsbankoti@gmail.com

**Abstract**

A study on North-South (N-S) asymmetry of different solar activity features (DSAF) such as solar proton events, solar active prominences, Hα flare index, soft X-ray flares, monthly mean sunspot area and monthly mean sunspot number were carried out from 1996-2008. It is found in our result that solar cycle 23 is magnetically weak compared to solar cycle 22. Study shows the Southern dominance of DSAF during the time period of study. During the rising phase of the cycle the numbers of DSAF approximately equal on the North and South Hemisphere. However, these activities tend to shift from Northern Hemisphere to Southern Hemisphere in between year 1998-1999. The statistical significance of the asymmetry time series using a $\chi^2$ - test of goodness of fit indicates that in most of the cases the asymmetry is highly significant, i.e., the asymmetry is a real feature in the N-S distribution of DSAF.

## 1. Introduction

The occurrence of various solar activity features shows non-uniformity over the solar disk. It has also been noticed that more activity features occur in one part of the solar Hemisphere than the other at one time. When these features are examined with respect to equator of the Sun, they are referred to as the North-South (N-S) asymmetry. The N-S asymmetry of several solar activity features viz relative sunspot numbers (SN), sunspot groups, sunspot areas (SA), flares, Hα flare indices (Q), solar active prominences (SAP), coronal mass ejections (CME), radio bursts, solar gamma ray bursts, coronal holes, faculae, green corona, corona's ionization temperatures and magnetic fluxes have been investigated by various authors (Maunder, 1904; Howard, 1974; Roy, 1977; Vizoso and Ballester, 1987; Bai, 1990; Oliver and Ballester, 1994; Joshi, 1995; Ataç and Özgüç, 1996; Verma, 2000; Temmer et al., 2001, 2002; Joshi and Joshi, 2004; Joshi and Pant, 2005; Atac and Özgüç, 2006; Gao, Li and Zhong, 2007; Chang, 2008; Li et al., 2009; Joshi et al., 2009).

Bell (1962) has found long term N-S asymmetry in the SA data for SC 8-18. Carbonell et al. (1993) presented a thorough study of the N-S asymmetry of daily SA, since these are good indicators of magnetic activity and found the N-S asymmetry is statistically highly significant using $\chi^2$ - test of goodness of fit. A detailed study has been carried out by Verma (2000) to decipher N-S asymmetry of SAP during 1957-1998. The existence of the real N-S asymmetry of SXR flare index (which is strengthened during solar minimum) during solar cycle 21, 22 and 23 has been analyzed (Joshi and Joshi, 2004). Joshi et al. (2005) also investigated the N-S asymmetry of Hα flare events during the solar cycle 23. Zaatri et al. (2006) studied the N-S asymmetry of meridional and zonal components of horizontal, solar subsurface flows during the years 2001–2004, which cover the declining phase of solar cycle 23. More recently, asymmetry of solar activity in cycle 23 has been studied by Li et al. (2009) using sunspot groups and SA from May 1996 to February 2007 and find the solar activity dominance in the Southern Hemisphere for cycle 23. Joshi et al. (2009) studied the asymmetry of SAP and groups of different limb and disk features of SAP for solar cycle 23 and presented a comparison among the solar cycles 21, 22 and 23.

In the past, the N-S asymmetries of several solar activity features on the entire solar disk and atmosphere have been studied by various researchers (Waldmeier, 1971; Verma, 1987; 1992; 1993; Ballester et al., 2005). Waldmeier (1971) studied asymmetry of most outstanding feature of solar activity in the decade 1959 -1969 on the two Hemispheres and found that on the Northern Hemisphere spots, faculae and prominences were more numerous and the white light corona was brighter than on the Southern Hemisphere. He also found that the green coronal line was brighter on the Northern Hemisphere, but the intensity of the red line was asymmetric in the opposite sense. Further, a comprehensive study was carried out by Verma (1987) on N-S asymmetry for major flares (solar cycles 19 and 20), type II radio bursts (solar cycles 19, 20 and 21), white light flares (solar cycles 19, 20 and 21), gamma ray bursts, hard X-ray bursts and CME (SC 21) and the results were compared with the found asymmetry in favor of the Northern Hemisphere during solar cycle 19 and 20 in favor of the Southern Hemisphere during solar cycle 21. Verma (1992) also predicted that the N-S asymmetry in solar cycles 22, 23 and 24 may be Southern dominated which will shift to Northern Hemisphere in solar cycle 25. Verma (1993) studied various solar phenomena occurring in both Northern and Southern Hemisphere of the Sun for solar cycles 8-22, calculated the N-S asymmetry indices for several solar phenomena and plotted them against the number of solar cycles. Similarly, Ballester et al. (2005) have studied N-S asymmetry using different phenomena of solar activity. Summaries of the studies of Hemispherical asymmetries of solar activity have been included in the work of Vizoso and Ballester, 1990; Li et al., 1998 and Li et al., 2002.

The aim of our work is to make a detailed study of N-S asymmetry of DSAF (SN, SA, SXR, Hα flare index, SAP, SPE) from 1996 to 2008 (solar cycle 23). Section 2 shows the observational data and statistical analysis. Section 3 represents a brief description of N-S asymmetry of DSAF during the considered time period. Section 4 contains discussions and results.

## 2. Observational Data and Statistical Analysis:

The different solar activity features analyzed in our study have been downloaded form following various web sites:

- The monthly North and South SNs (1996-2008), from *ftp://ftp.ngdc.noaa.gov*/STP/SOLAR_DATA/SUNSPOT_NUMBERS/RNRS with 682 data points.
- The monthly North and South number of SAPs (1996-2008), obtained from *ftp://ftp.ngdc.noaa.gov*/STP/SOLAR_DATA/SOLAR_FILAMENTS with 8539 data points. After excluding the events which lie on equator, a total number of 8426 events have been selected to study latitude and longitude distribution.

- The monthly North and South SXR flares (1996-2008), detected by the GOES satellites, which can be downloaded from *ftp://ftp.ngdc.noaa.gov*/STP/SOLAR_DATA/SOLAR_F*LARE/XRAY_FLARES* with 23335 data points. Out of this 12127 flare events for which heliographic coordinates are given, are selected.
- The monthly number of North and South flare index (1996-2007) compiled in the Kandilli Observatory, from *ftp://ftp.ngdc.noaa.gov*/STP/SOLAR_DATA/SOLAR_*FLARES/INDEX.*
- The monthly North and South SA (1996-2008) compiled by D.Hathaway, from *http://solarscience.msfc.nasa.gov/greenwch.shtml*.
- The North and South monthly number of solar proton events (SPE) (1996-2008), from *http://www.swpc.noaa.gov/ftpdir/indices/SPE.txt* with 72 data points.

We compare the amplitudes of solar cycle 22 and 23 by using DSAF. In all activity indices the amplitudes of solar cycle 23 are distinctively weaker than the solar cycle 22, as can be seen from Fig. 1 and Table 1. First column of Table 1 represents the DSAF for cycle 22 and 23. 2$^{nd}$, 3$^{rd}$, 4$^{th}$ and 5$^{th}$ columns represent the average values of the DSAF as well as their standard deviations within brackets during the solar maximum and minimum phases of the solar cycles 22 and 23. The last two columns illustrate the variability of DSAF during rising phase and decay phase between the above two successive cycles.

The N-S asymmetry of solar activity phenomena has been calculated using the formula:

$$A_{NS} = \frac{N-S}{N+S} \tag{1}$$

Here, $A_{NS}$ is the N-S asymmetry index, N is the number of solar activity phenomena in Northern Hemisphere and $S$ is the number of solar activity phenomena in Southern Hemisphere. Thus, if $A_{NS} > 0$, the activity in the Northern Hemisphere dominates and if $A_{NS} < 0$, the reverse is true. To verify the reliability of the observed N-S asymmetry values, a $\chi^2$ - test is applied with Equation (2) given below.

$$\chi_{NS} = \frac{2(N+S)}{\sqrt{(N+S)}} = \frac{\sqrt{2}A_{NS}}{\Delta A_{NS}} \tag{2}$$

If $A_{NS} < \Delta A_{NS}$, $\Delta A_{NS} \le A_{NS} < 2\Delta A_{NS}$ and $A_{NS} \ge 2\Delta A_{NS}$, the probability that N-S asymmetry exceeds the dispersion value is $p < 84\%$, $84\% \le p < 99.5\%$ and $p \ge 99.5\%$ respectively. Here the first, second and third limits imply statistically insignificant, significant and highly significant values respectively (Joshi and Joshi, 2004; Joshi et al., 2009).

## 3. N-S Asymmetry of Sunspot Numbers, Sunspot Area, SXR Flares, Hα Flare Index, Solar Active Prominences and Solar Proton Events

To investigate the behavior in variation of DSAF in the North and South Hemispheres, activities were counted for each Hemisphere separately and are shown in Table 2 & Figure 2. In Table 2, first and second column represent DSAF and total number of events reported from different ground and space based observatories and satellites respectively. Column 3rd represents events for which the heliographic coordinates are given and column 4th and 5th represent the total number of events in Northern and Southern Hemisphere respectively during the considered time span. All parameters shown in Fig 2 show nearly the same monthly variation in the North-South Hemisphere except SAP. Further, SAP shows an abnormal behavior as it is active from 1996-1998. On the other hand, for the same period, other activities are almost dormant. For more clear results smoothened monthly means of the DSAF indices in the Northern and Southern Hemispheres are shown in Fig. 3 (left panel). In the temporal variation of SA, SXR flares, SAP and SPE the strength of the activity of two Hemispheres is different. The strength dominates in Northern Hemisphere for flare index and SAP whereas it dominates in Southern Hemisphere for SA, SXR and SPE. Additional plot of the smoothed asymmetry for the monthly mean is given for each solar active feature in Fig.3 (right panel). It is clear that the dominance of activity at the beginning of the cycle started to alternate between the two Hemispheres during the rising phase (0 – 48 months). However during the maximum phase the N-S asymmetry dropped nearly to zero and remained very low (48 – 96 months) and after 96 months of the considered time it is in favor of Southern Hemisphere.

All these results are confirmed by Fig. 4 in which we have shown the cumulative counts of the monthly DSAF for the Northern (black line) and Southern Hemispheres (red line) respectively. The vertical spacing between the two lines is a measure of the Northern/Southern excess of DSAF at that time which shows Southern dominance of SN, SA, SXR, Q and SAP excluding SPE. In order to characterize the N-S asymmetry of DSAF, the asymmetry values of the monthly DSAF numbers in the Northern and Southern hemispheres were calculated, and fitted into a straight line (Fig. 5). Here, interestingly, we notice that, during considered time period, the behavior of asymmetry changes from North to South hemisphere in the year 1998-99 for SN, SA, Q and SXR whereas asymmetry of SAP changes from North to South in the year 1997. From Fig. 5 it is clearly evident that the slopes are negative for each activity which again supports our previous observation that the activities are Southern Hemisphere dominant. To understand the behavior of DSAF with respect to SN, correlations were obtained between DSAF and SN (Fig. 6) and following points were noticed; (i) Higher correlation (positive) for SA and Q (ii) Moderate correlation (positive) for SXR and SPE and (iii) Poor correlation (negative) for SAP. In Fig. 7 we have plotted the annual asymmetry index of DSAF. Yearly N-S asymmetry of DSAF where highly significant, significant and insignificant values are marked with filled square (■), circle (●) and triangle (▲) respectively, using $\chi^2$ - test (equation 2). Significance level of N-S asymmetry is also represented in Table 3. SXR flare, SA and SAP have larger number of highly significant values as compared to SN and Q.

## 4. Discussion and Results:

We have presented a comparative study of activity between cycles 22, 23 using DSAF. In Table 1 we have represented the minimum to maximum variability of the DSAF during cycle 22 and 23. In column 5 of this table all the features except SAP show values greater than 1, which means that the activity in rising phase of solar cycle 22 is higher than the solar cycle 23. The SAP shows negative value because there are very less number of SAP observed in maximum phase compared to minimum phase of solar cycle 23. In column 6 all the features show values greater than 1, which means that the activity in solar cycle 22 is higher than solar cycle 23. The number of SAP events is 27.47 times larger in decay phase of cycle 22 than solar cycle 23. During the time of increased amplitude of solar activity we would expect short-period cycles and during the times of decreased amplitude of solar activity we would expect loner-period cycles as in cycle 23 (Hathaway et al., 2002). There is enough evidence to support that the chosen cycle 23 is magnetically weak. This peculiarity is clearly evident from comparison of DSAF during cycle 22 and 23 (Fig. 1). Even-Odd (or Gnevyshev, G-O) effect (Gnevyshev and Ohl, 1948) is seen in the monthly averaged or smoothed wolf sunspot numbers (1749-2003) where the odd-numbered cycle amplitudes are compared with the amplitudes of the preceding even-numbered cycles. Our results show the violation of G-O rule in cycle 23 for DSAF.

We have also presented the N-S asymmetry of DSAF for solar cycle 23 and the obtained results are discussed below. Earlier studies by Verma (1992, 93); Ataç and Özgüç, (1996, 2006) and Li et al. (2002) gave the dominant Hemisphere of solar activity in each of solar cycles 12 to 22 and inferred that a Southern dominance of solar activity should occur in solar cycle 23. Recently Li et al. (2009) and Joshi et al. (2009) found out the Southern dominances of SA, grouped SN and different groups of SAP respectively. Our results support the predictions. This finding is further confirmed in the present study by investigation of the cumulative counts of DSAF number for the Northern and Southern Hemispheres during considered time. It is found that all the features except SPE and Q show a small spacing during the rising and maximum phase whereas a considerable spacing (Southern dominances) between the two cumulative lines during the end phase of solar cycle 23 (Fig. 4). Vizoso and Ballester, 1990; Ataç and Özgüç, 1996; Li et al., 2002 and Joshi and Joshi, 2004; found regularity in their studies of different activity parameters that for each four cycles the slope of the fitting straight lines of the N-S asymmetry values changes its sign. The solar activity of cycle 23 is very important for demonstrating the existence of these regularities. We have also demonstrated the Southern dominance of solar activity in cycle 23 with the help of linear fitting of monthly asymmetry index values of DSAF (see Fig. 5), which is similar to solar cycles 20, 21 and 22 as the obtained slopes are negative. Yearly N-S asymmetry of DSAF shows similar type of variation during cycle 23 (Fig. 7). In our study, we observe the DSAF show periodic behavior form 1996 to 2001. N-S asymmetry for SA, SXR, Q and SAP favors Southern Hemisphere in the year 1996. In 1997, 1999 and 2000 all the activity asymmetries move to the Northern Hemisphere. These shift to South again in 1998 and beyond 2001, where these strongly favor the Southern Hemisphere. It also has been found that the variation of N-S asymmetry of SPE has an alternating nature during whole cycle. Temmer et al. (2001) and Joshi et al. (2009) analyzed the N-S asymmetry of Hα flares and SAP and found high statistical significance of N-S asymmetry. We have also found highly significant values of all DSAF (Table 3), which is in agreement with previous studies (Temmer et al., 2001; Joshi and pant, 2005).

The physical interpretation of the asymmetry in the solar activity has not been given clearly so far. However, Hathaway (1996), argued that the fluctuations in the meridian flow, believed to be a product of turbulent convection and variations in the gradient of the rotation rate, contributed to the cycle amplitude variations. The Sun's magnetic activity is generally believed to be supplied by a hydro magnetic dynamo operating either in or at the base of the solar convective zone (Pulkkine et al. 1999). Zhao and Kosovichev (2004) have found out that meridian flows vary with the location of the solar activity belts. By applying time-distance helioseismology measurements, it is demonstrated that the activity belts which were located towards the pole side of the faster zonal bands migrated together towards the solar equator around the maximum phase of the solar cycle 23 and their synoptic maps show unsymmetrical zonal flows in the Northern and Southern Hemispheres (Haber et al. 2002). Zaatri et al. (2006) also found the zonal flow is larger in the Southern Hemisphere than the Northern one, and N-S asymmetry increases with depth. These explanations are only suggestive and more observations of local helioseismology are needed to understand the N-S asymmetry of DSAF.

**Acknowledgement**

Authors, NSB and NCJ are thankful to UGC, New Delhi, India for financial assistance under RFSMS (Research Fellowship in Science for meritorious students) scheme.

## Tables

**Table 1**: Rising and decay phase variability of DSAF during solar cycle 22 and 23.

| DSAF: Cycles 22 and 23 | 1986 Minimum | 1989-1991 Maximum | 1996 Minimum | 2000-2002 Maximum | 2008 Minimum | Rising phase cycle22/ cycle23 | Decay phase cycle22/ cycle23 |
|---|---|---|---|---|---|---|---|
| **Sunspot Number** | **13.4** (9.9) | **148.6** (23.2) | **8.6** (5.3) | **111.5** (19.4) | **2.9** (2.4) | 1.31 | 1.29 |
| **Sunspot Area** | **124.7** (134.7) | **2366.0** (654.8) | **81.9** (80.1) | **1716.8** (562.1) | **22.78** (39.6) | 1.37 | 1.35 |
| **Flare Index** | **1.2** (1.4) | **14.9** (4.6) | **0.4** (0.4) | **6.3** (4.3) | **0.46** (0.7) | 2.32 | 2.48 |
| **Hα flare** | **730** (54.6) | **20816** (117.8) | **280** (18.1) | **11294** (146.6) | **125** (30.6) | 1.82 | 1.84 |
| **x-ray flare** | **916** (62.8) | **8538** (58.5) | **515** (30.1) | **8085** (53.6) | **86** (10.2) | 1.01 | 1.003 |
| **Solar active Prominence** | **3095** (127.1) | **33891** (130.28) | **2213** (63.8) | **1163** (24.7) | **10** (1.2) | -29.33 | 27.47 |

**Table 2**: Total number of DSAF in Northern and Southern Hemisphere during solar cycle 23.

| Solar Active Feature | Total Number | Total Number (N+S) | Total North Number (%) | Total South Number (%) | Dominant Hemisphere |
|---|---|---|---|---|---|
| SAP | 8539 | 8426 | 4018 (47.69%) | 4408 (52.31%) | S |
| SXR | 23,335 | 12,127 | 5131 (42.31%) | 6996 (56.69%) | S |
| SPE | 72 | 71 | 36 (50.7%) | 35 (49.3%) | - |
| SN (monthly mean) | - | 682 | 322.1 (47.22%) | 359.9 (52.78%) | S |
| SA (monthly mean) | - | 10091.16 | 4690.1 (46.48%) | 5401.06 (53.52%) | S |
| Q (yearly) | - | 443.53 | 212.95 (50.26%) | 230.58 (49.74%) | - |

**Table 3**: Significance level of DSAF during solar cycle 23 using $\chi^2$-test.

| Solar Active Feature | Highly Significant | Significant | Insignificant |
|---|---|---|---|
| SAP | 8 | 1 | 4 |
| SXR | 11 | 2 | 0 |
| SPE | 4 | 5 | 4 |
| SN (monthly mean) | 5 | 3 | 5 |
| SA (monthly mean) | 10 | 2 | 1 |
| Q (yearly) | 6 | 2 | 4 |

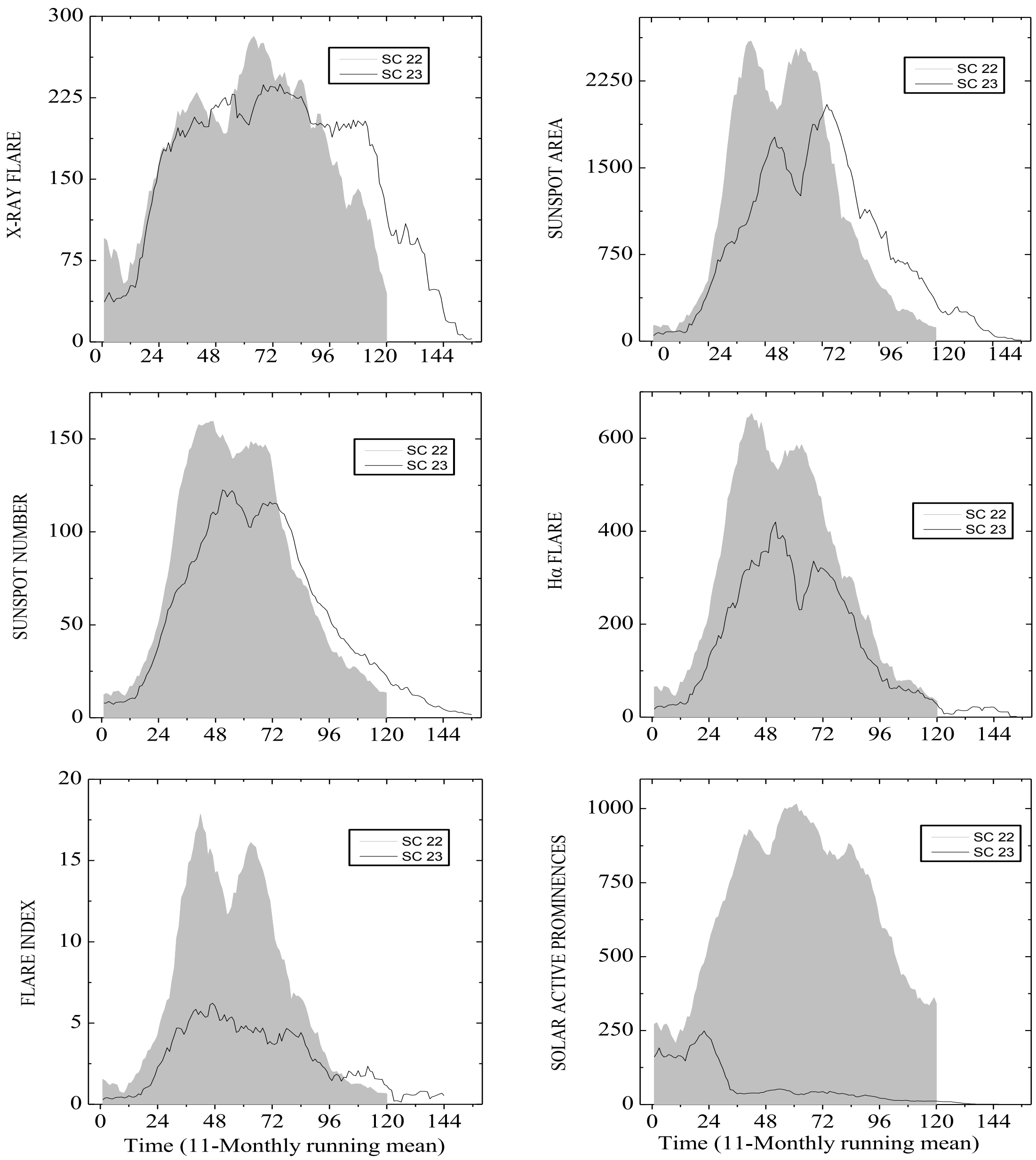

**Figure 1**: Comparison of the monthly mean of the SXR, SA, SN, Hα flare, Q and SAP for the solar cycle 22 and solar cycle 23.

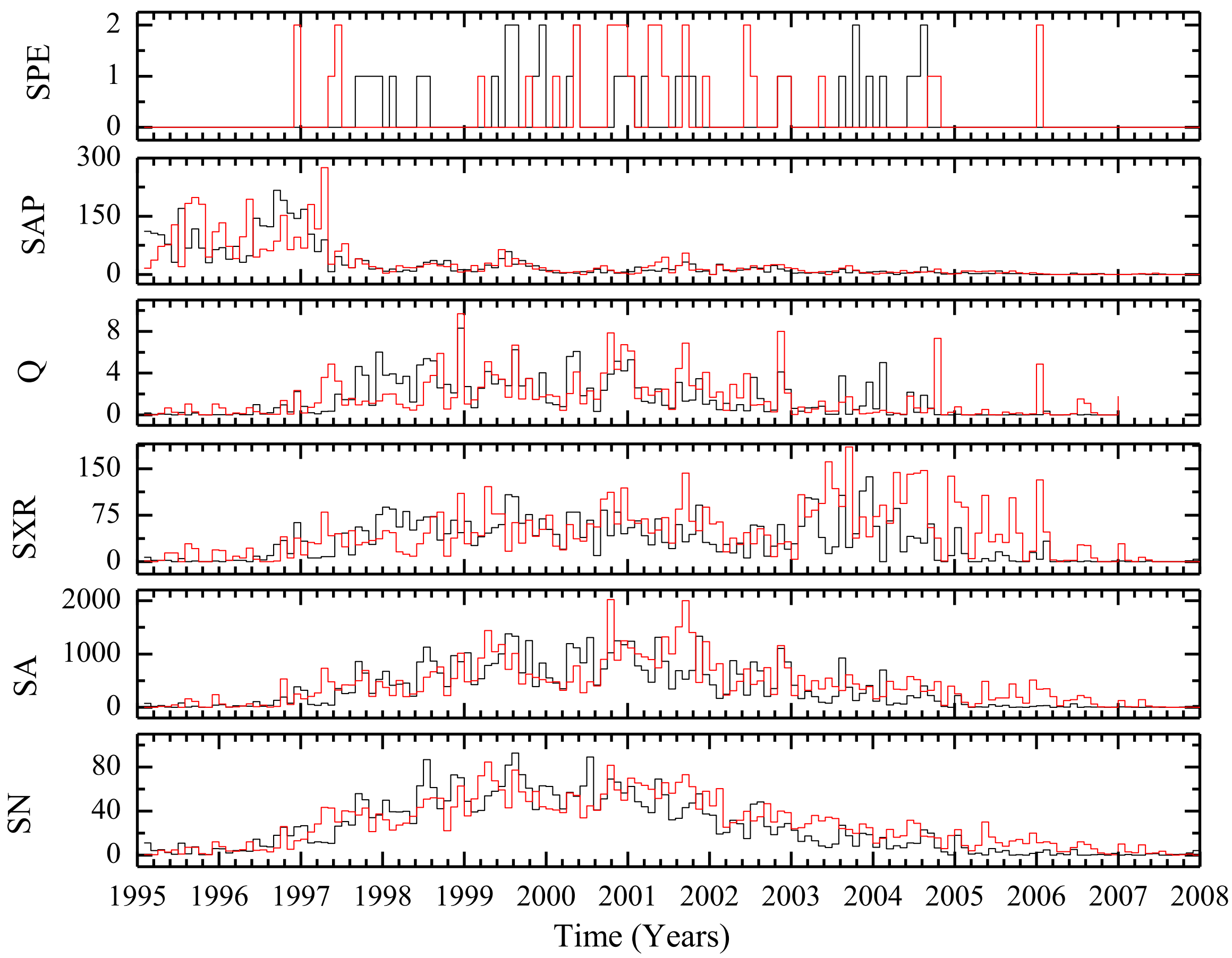

**Figure 2**: Monthly number of solar proton events (SPE), solar active prominences (SAP), Flare index (Q), Hα flare, soft X-ray flare (SXR), monthly mean sunspot area (SA) and monthly mean sunspot number (SN) respectively in the Northern (the black line) and Southern (the red line) Hemispheres from 1996-2008 (from top to bottom ).

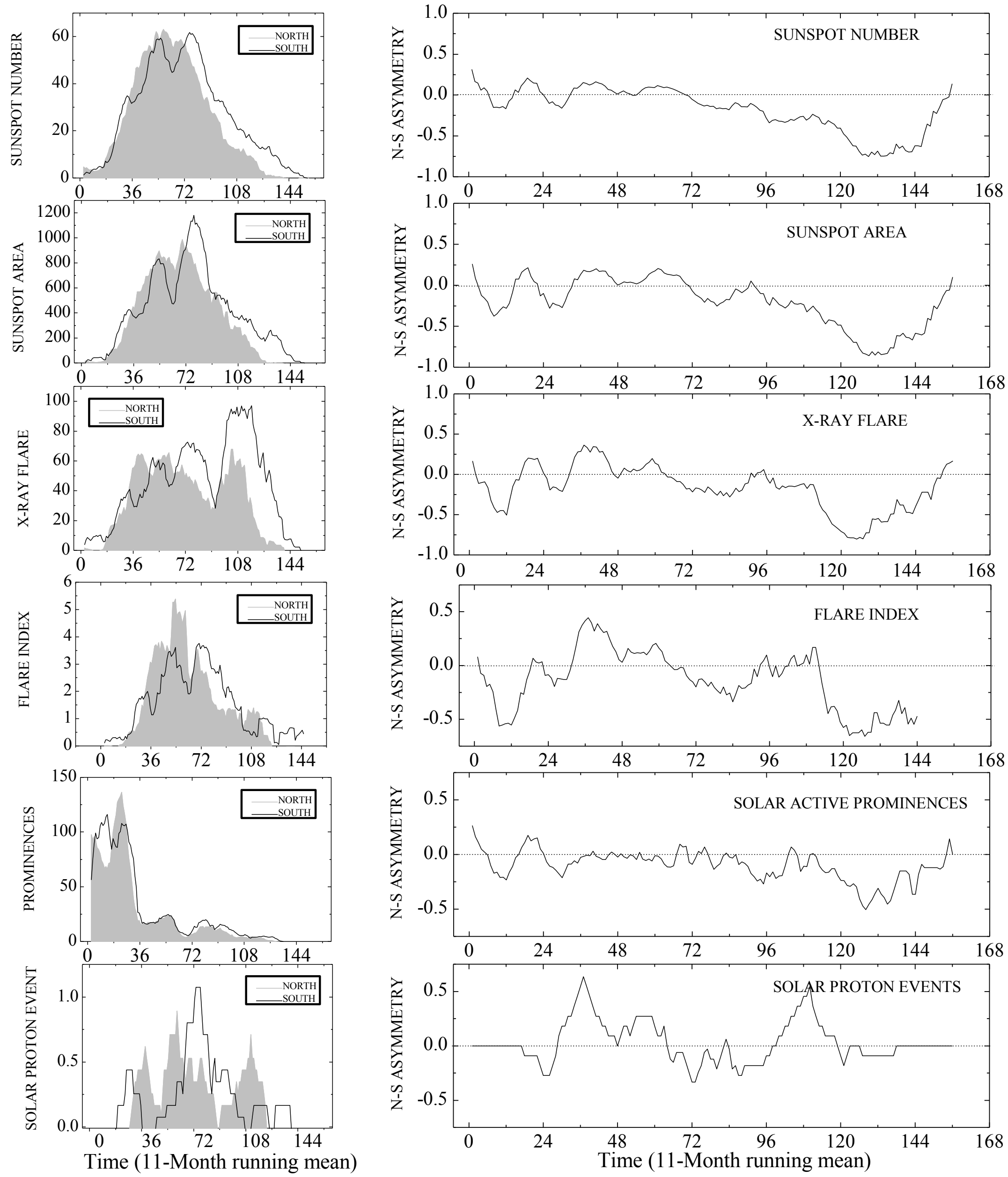

**Figure 3.** Monthly plots of N-S asymmetry of the SN, SA, SXR, Q, SAP and SPE and Gnevyshev gap time occurrence.

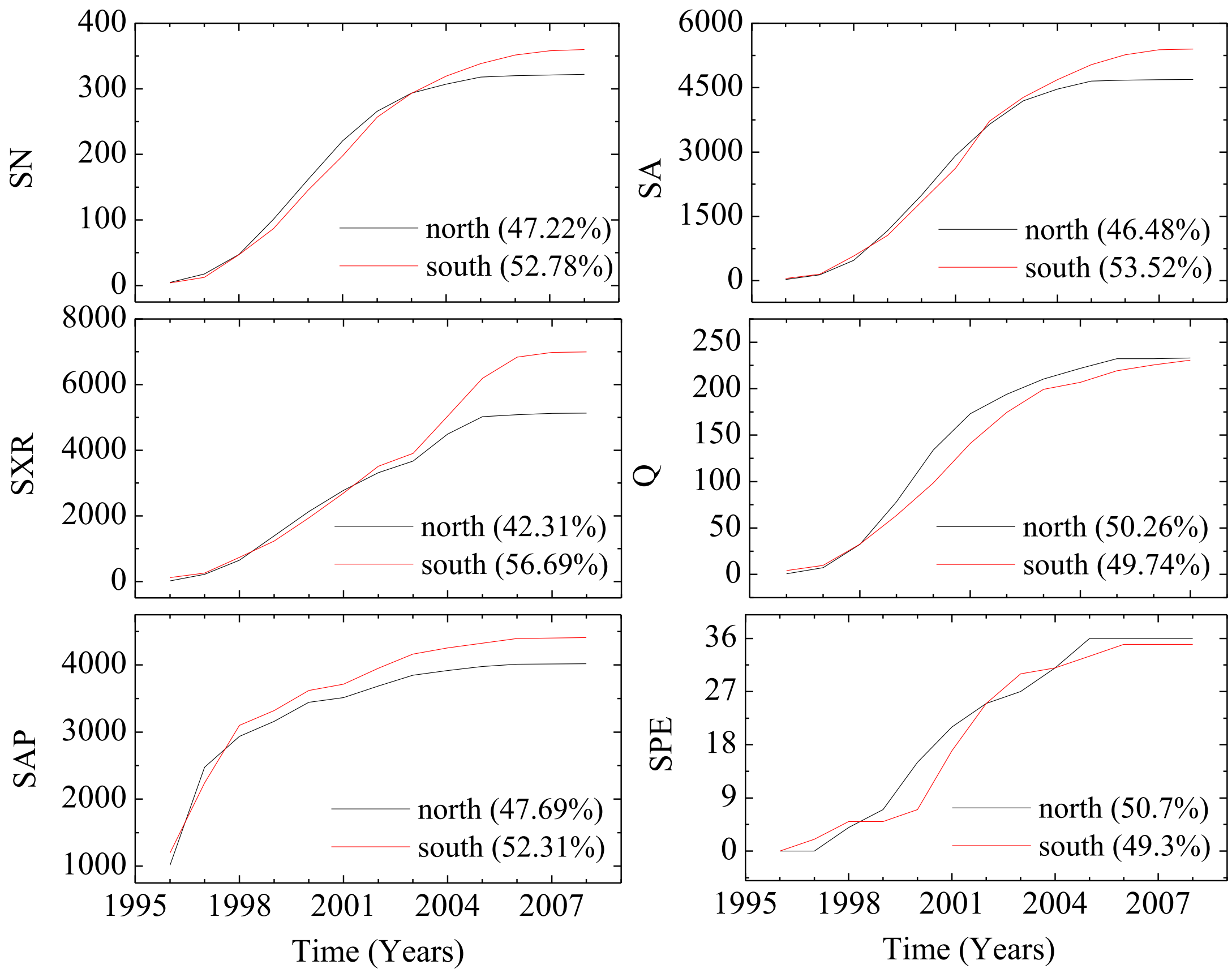

**Figure 4:** Cumulative monthly mean sunspot number (SN), monthly mean sunspot area (SA) number of soft X-ray flare (SXR), Hα flare index (Q), solar active prominences (SAP) and solar proton events (SPE) respectively in the Northern Hemisphere (the black lines) and the Southern Hemisphere (the red lines) from 1996-2008.

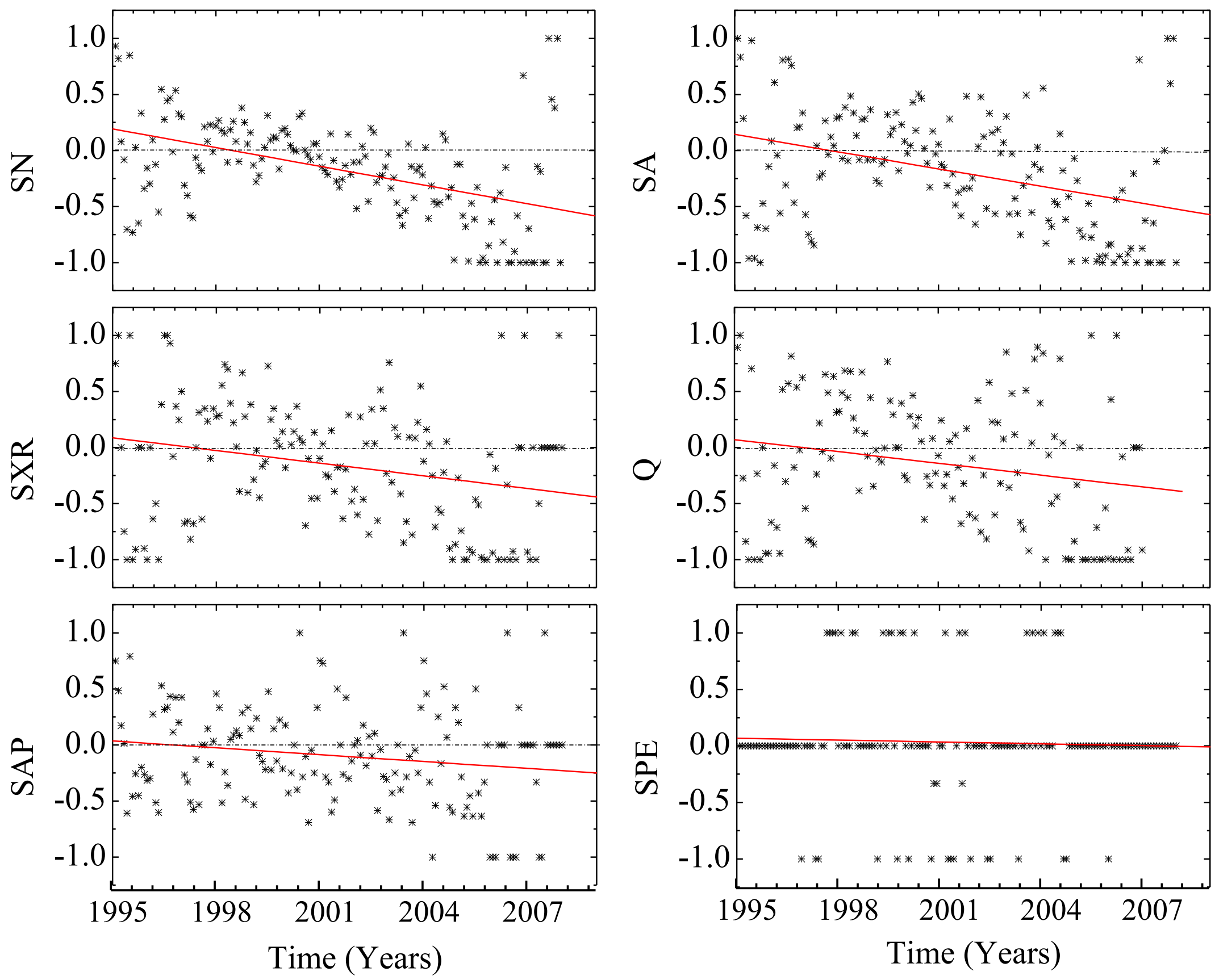

**Figure 5**: Fit of a regression straight line to the monthly values of the North-South asymmetry of monthly mean sunspot number (SN), monthly mean sunspot area (SA) number of soft X-ray flare (SXR), Hα flare index (Q), solar active prominences (SAP) and solar proton events (SPE)respectively from 1996-2008.

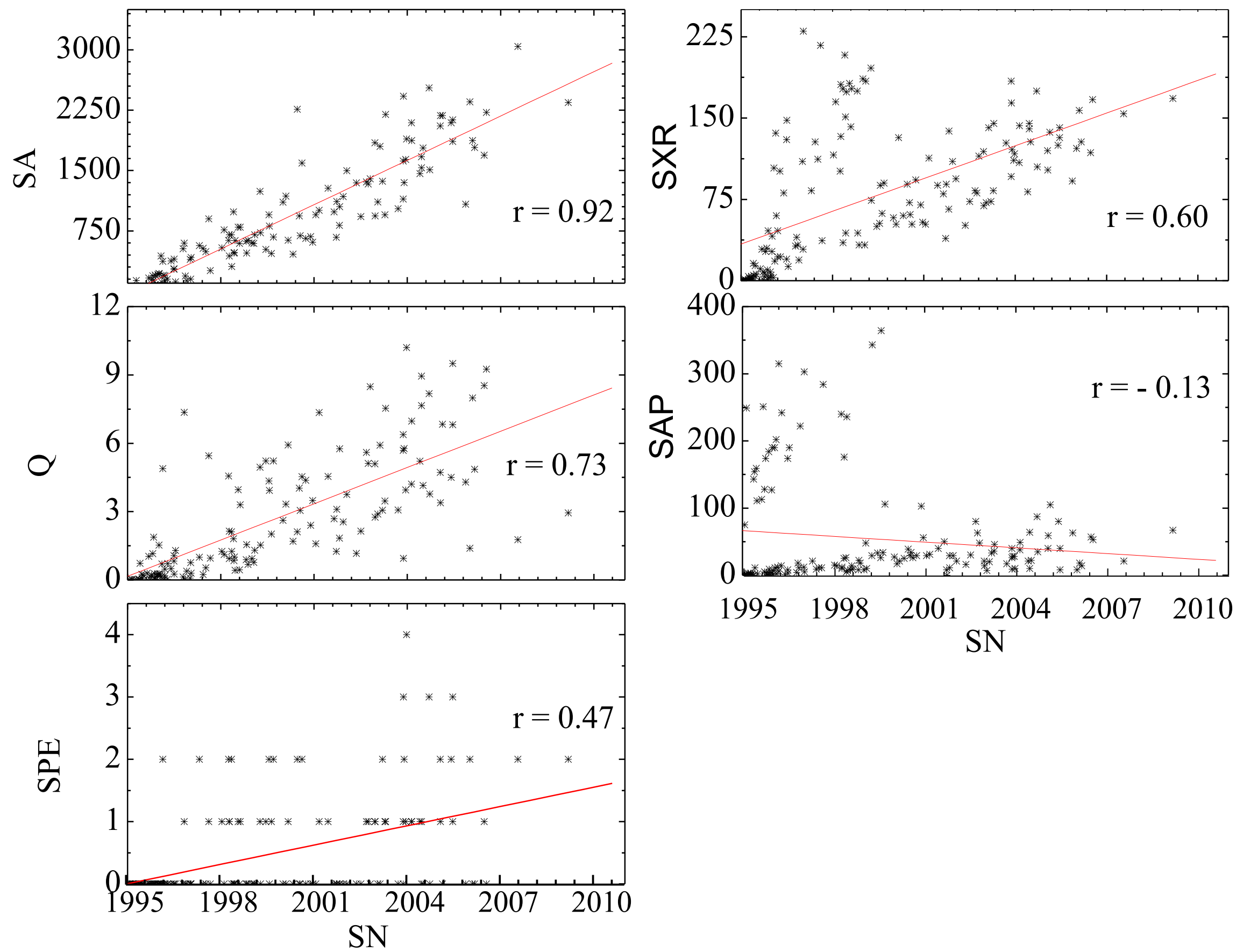

**Figure 6**: Correlation plot of the monthly mean sunspot area (SA), soft X-ray flare (SXR), Hα flare index (Q), solar active prominences (SAP) and solar proton events (SPE) with monthly mean sunspot number (SN); r indicates the correlation coefficient.

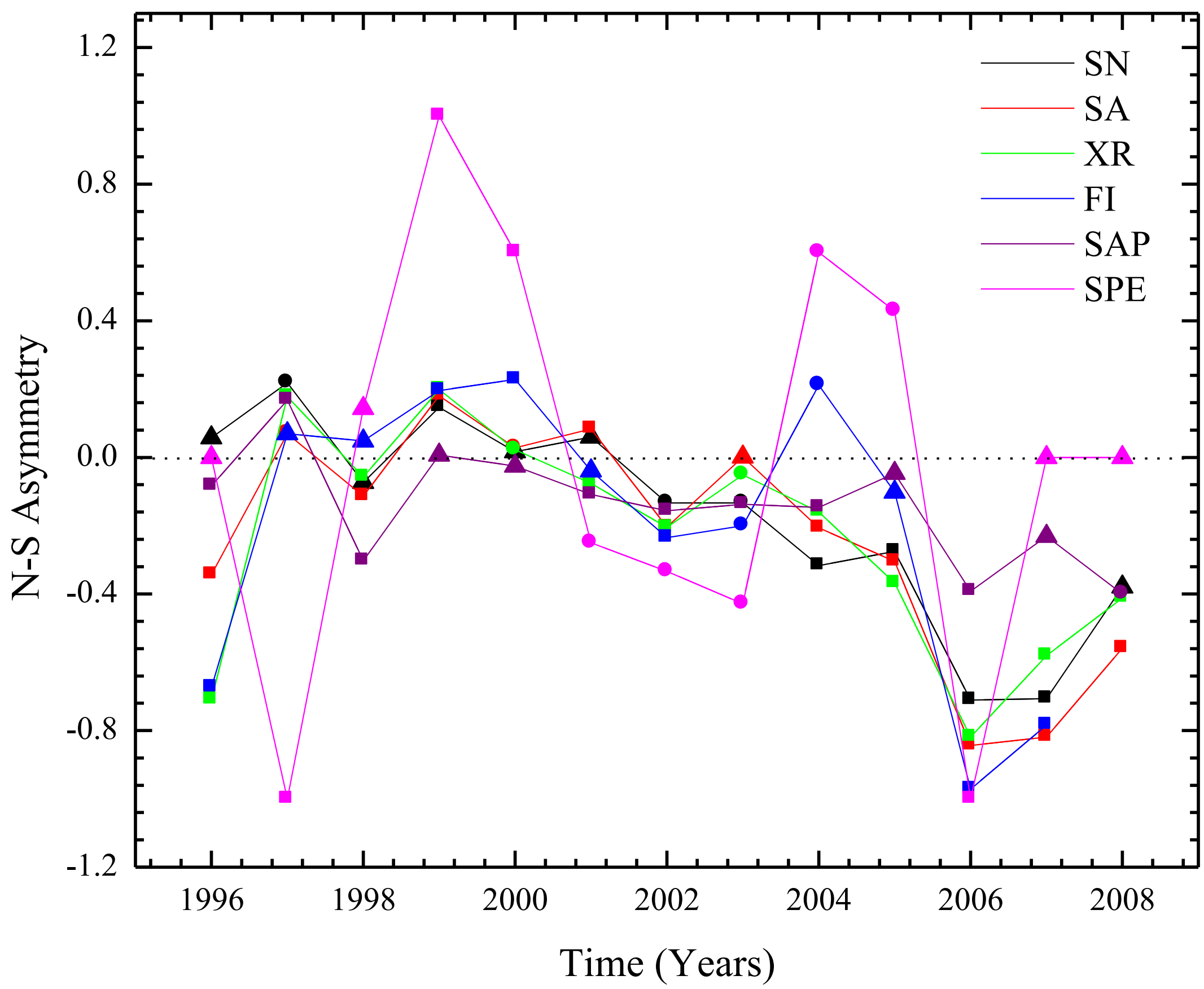

**Figure 7:** Plot of N-S asymmetry of Various Solar Active Phenomena versus years. Highly significant, significant and insignificant values marked with filled square (■), circle (●) and triangle (▲) respectively from 1996-2008.

--------------------